\RequirePackage{fix-cm}
\documentclass[onecolumn,epjc3]{svjour3}  
\smartqed  %
\RequirePackage{graphicx}

\usepackage{amsmath}
\usepackage{amssymb}
\usepackage{epstopdf}
\usepackage{color}
\usepackage{mathptmx}
\DeclareGraphicsExtensions{.eps,.pdf,.png,.gif,.jpg,.mht}

\journalname{Jinst}

\date{\today}
\begin{document}

\title{ Timing Performance of a  Double Layer Diamond Detector}

\author{
M.~Berretti\thanksref{a1} \and
E.~Bossini\thanksref{5,4} \and
M.~Bozzo\thanksref{3} \and
V.~Georgiev\thanksref{2} \and
T.~Isidori\thanksref{4} \and
R.~Linhart\thanksref{2} \and
N.~Turini\thanksref{6}
}

\institute{
University of Helsinki and Helsinki Institute of Physics, P.O. Box 64, FI-00014, Helsinki, Finland. \label{a1} \and
Universit\`a degli studi di Pisa and INFN di Pisa, Largo B. Pontecorvo 3, 56127 Pisa, Italy. \label{4} \and
Centro Studio e Ricerche Enrico Fermi Piazza del Viminale 1 - 00184 Roma (Italy). \label{5}\and
INFN Sezione di Genova, Via Dodecaneso 33, 16136 Genova, Italy. \label{3} \and
University of West Bohemia, Pilsen, Univerzitni 8, PILSEN 30614 (Czech Republic). \label{2} \and
Universit\`{a} degli Studi di Siena and Gruppo Collegato INFN di Siena, Via Roma 56, 53100 Siena, Italy. \label{6}
}

%
%

%
\thankstext{e1}{Corresponding author's e-mail: mirko.berretti@cern.ch}
\maketitle
\abstract{In order to improve the time precision of detectors based on diamonds sensors we have built a detector with two scCVD layers connected in parallel to the same amplifier. 
This note describes the design and the first measurements of such a prototype performed on a particle beam at CERN.
With this different configuration we have obtained an improvement on the timing precision of a factor of 1.6-1.7 with respect to a single layer scCVD diamond detector.

\PACS{
     } %
}

\section{Overview}
\label{sec:over}
Diamond detectors are presently used extensively in many LHC experiments (\cite{Gorisek:2007cv}, \cite{Cindro:2008zz}, \cite{Bartz:2009zza}, \cite{Domke:2008dma}) and in ion facilities (GSI) to monitor beam condition and rates.

The TOTEM collaboration exploits the diamond detector good timing capabilities to measure with the utmost precision the arrival time of the leading proton in a scattering event and has recently installed in the LHC a set of timing detectors on each side of the interaction point IP5.
The detectors are installed in the very small space inside a movable section of the vacuum pipe of the CERN LHC, the Roman Pot (RP); details are described in~\cite{timing-nov-15}.

As expected tests shows that improving the S/N of the amplifier is the key element to obtain a better time precision and hence in the case of TOTEM a better longitudinal precision in the determination of the interaction vertex, given by $z= c\, \Delta T/2$ where $\Delta T$ is the difference of the arrival time of the protons in the left and right RP, $c$ is the velocity of light.

For a given amplifier to have larger signal one should increase the primary charge collected by the scCVD diamonds. For a larger charge collection one may increase the detector thickness, at the expense of a longer collection time.
In this article we will instead use two separated scCVD diamonds in parallel to the same amplifier channel in order to double the collected charge while keeping the collection time constant.
\section{Principle}
\label{sect:princ}
In 2004, in order to improve the signal amplitude of a pCVD beam monitor detector, H. Pernegger tested a configuration of sensors such that the charge released in two sensors traversed simultaneously by the same charged particle is sent to the same amplifier input: the ``double-diamond assembly'' (DD)~\cite{1546469}.
Later also Benardzick~\cite{benardzick:2007} and Cardarelli~\cite{Cardarelli:2014hfa} tried a similar configuration.

Applying this idea to the precise measurement of the time from the signals of n diamond pixels traversed by the same particle and connected in parallel to the input of the same amplifier should improve the measurement of  the time precision by a factor $\sqrt n$ if this is compared to the combined time precision from the signals of n pixels each measured independently with its amplification chain.

Note that  in the context of a time measurement the connection of two sensors in parallel is different and gives better results than simply doubling the thickness of the sensor: for the same ``volume'' of sensor two separate parallel sensors will provide on the average the same amount of charge to the input of the amplifier with approximately half the collection time.

The negligible leakage current of the single crystal diamond sensor with respect to the noise current of the amplifier is another decisive aspect towards the interest in a parallel connection of two diamonds. 

Besides if the space available is limited (usually as in our case the thickness per layer of detector is the thickness of the electronic board) it may be more interesting to increase the number of diamond sensors per detector layer, obtaining with the same amount of sensors and half the number of boards a better time precision  than the one that could be obtained from 2n independent diamond boards in the same space. 
This connection scheme bring also advantages on the overall system aspects, reducing the total number of channels to acquire for a certain overall thickness of the detectors and the the total amount of heat dissipated by the system.

Thanks to the small dielectric constant of the diamonds, the capacitance of pixels of area $1-2\,mm^2$ is only about 0.2 pF which is still smaller than the input capacitance of the preamplifier ($\approx 0.4 \,$pF see~\cite{timing-nov-15}) therefore more than one diamond sensor can be connected in parallel to one amplifier without degradation of the timing performance if, as mentioned before, the leakage current of the single crystal diamond sensor is negligible when compared to the noise current of the amplifier.

In conclusion more than one sensor can be connected in parallel to the same amplifier: since the collected charge is double on average and even if the rise-time of the signal is the same as for a single sensor, we expect an improvement of the time precision of almost a factor 2  with respect to the one obtained from  a single measurement.

\section{Description of the Detector}

\begin{figure}[hb]
\centering
 \begin{center}
\includegraphics[height=.45\linewidth, width=.99\linewidth]{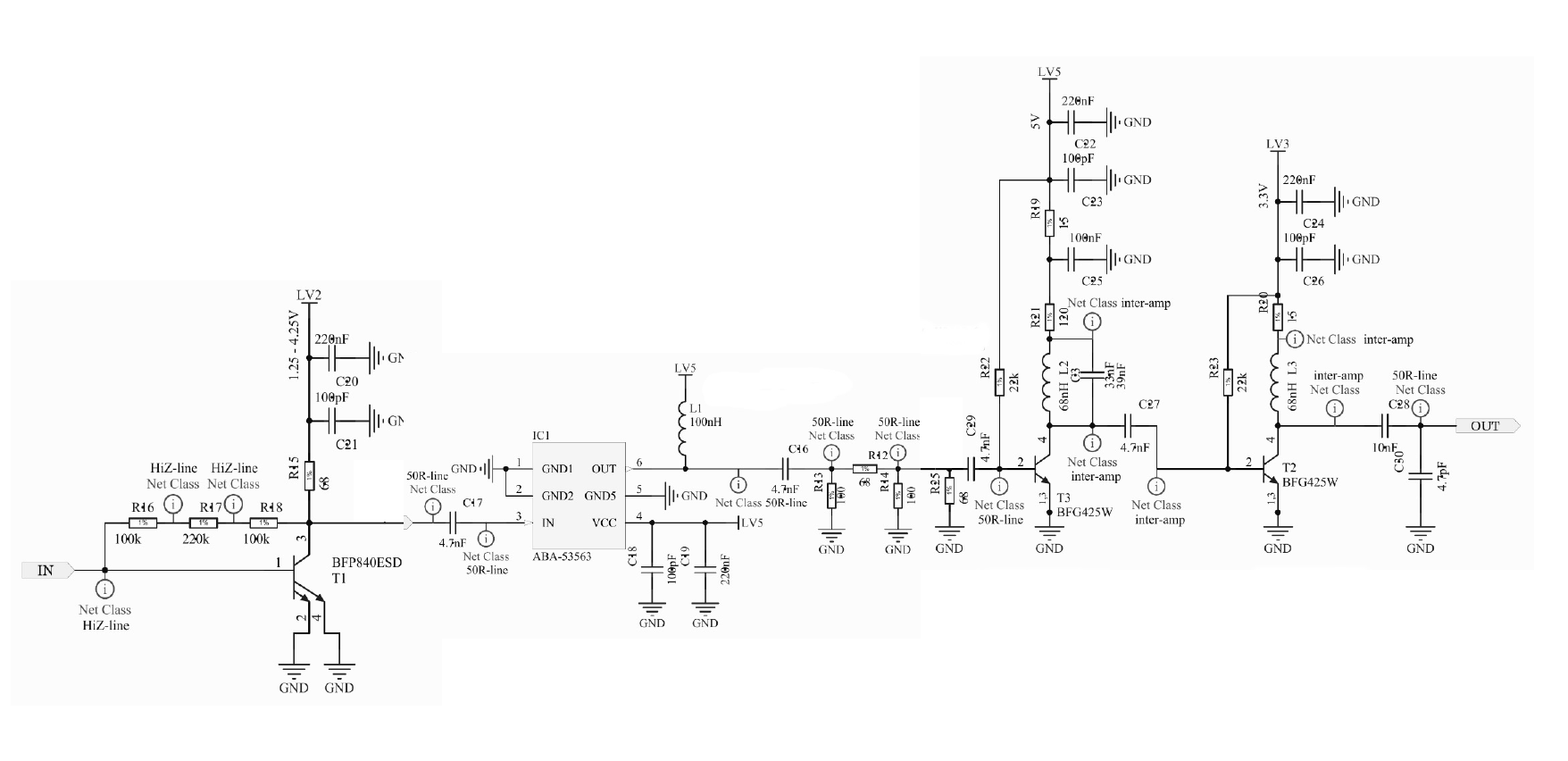}
\caption{Schematic of the TOTEM amplifier used for the measurements.}
\label{fig:amplifier-all}
\end{center}
\end{figure}

Having developed a new fast and performing electronics to measure timing with a precision down to less than 100 ps we decided to test it with a double-diamond assembly; the amplifier is described in details elsewhere~\cite{timing-nov-15} and its schematics reproduced in Figure~\ref{fig:amplifier-all}.
Diamond single crystal sensors are now easily available with a thickness of 500 $\mu$m, larger thickness with good quality crystals are difficult to obtain and reflects also on the cost per unit volume.

Two scCVD diamonds sensors of $4.5 \times 4.5\,$mm$^2$ each 500$\mu$m thick are precisely aligned and glued with conductive epoxy to the HV contact present on each side of a slightly modified hybrid board developed by TOTEM.  
The two diamonds aligned together for this measurement have a pixel structure: the pixel used for the measurement has $0.55 \times 4.2\, mm^2$ and for 2 pixels in parallel the area and capacity is the double.
The double diamond signal is compared with a signal obtained from a single scCVD diamond of similar area (pixel area $1.02 \times 4.2 \,{\rm mm}^2$), and therefore its capacitance is within 10\% equal to the one of the double arrangement.
The single and double diamond are glued on the same board. 
The signal electrodes from a double diamond are bonded with standard $25\,\mu m$ diameter Al wire to the input of the same amplifier, which will now sees summed the charge released in the two sensors.

Table~\ref{table:electronics} gives the main parameters of the electronics~\cite{timing-nov-15}  and Figure~\ref{fig:amplifier-all} is the schematic diagram of the same.

\begin{table}[!]
\centering
\caption{Characteristics of the TOTEM amplifier.}
\begin{tabular}{lcc}

\hline
Number of stages  & \multicolumn{2}{l}{ 4 } \\
\hline
& \multicolumn{2}{l}{transconductance amplifier} \\
 &  \multicolumn{2}{l}{flat freq response amplifier (ABA)} \\
 & \multicolumn{2}{l}{10 dB attenuator} \\
 &\multicolumn{2}{l}{ booster-shaper} \\
 \hline
 & at DC & at 200 MHz \\
\hline
Input impedance & 10 kOhm  &1.5 kOhm \\
Current gain &  & 93 dB \\
Current gain 1$^{st}$ stage & 45 dB & 30 dB \\
Power gain 2$^{nd}$ stage &  \multicolumn{2}{c}{21 dB} \\
Attenuation 3$^{rd}$ stage &  \multicolumn{2}{c}{10 dB} \\
Power gain 4$^{th}$ stage &  & 52 dB \\
Power dissipated & 0.3 W/ch \\

\hline
\end{tabular}

\label{table:electronics}
\end{table}

\section{Measurements}

The measurement was performed in a 180 GeV pion beam  at the CERN SpS. 

\begin{figure}[h]
 \begin{center}
\centering
\includegraphics[width=.4\linewidth]{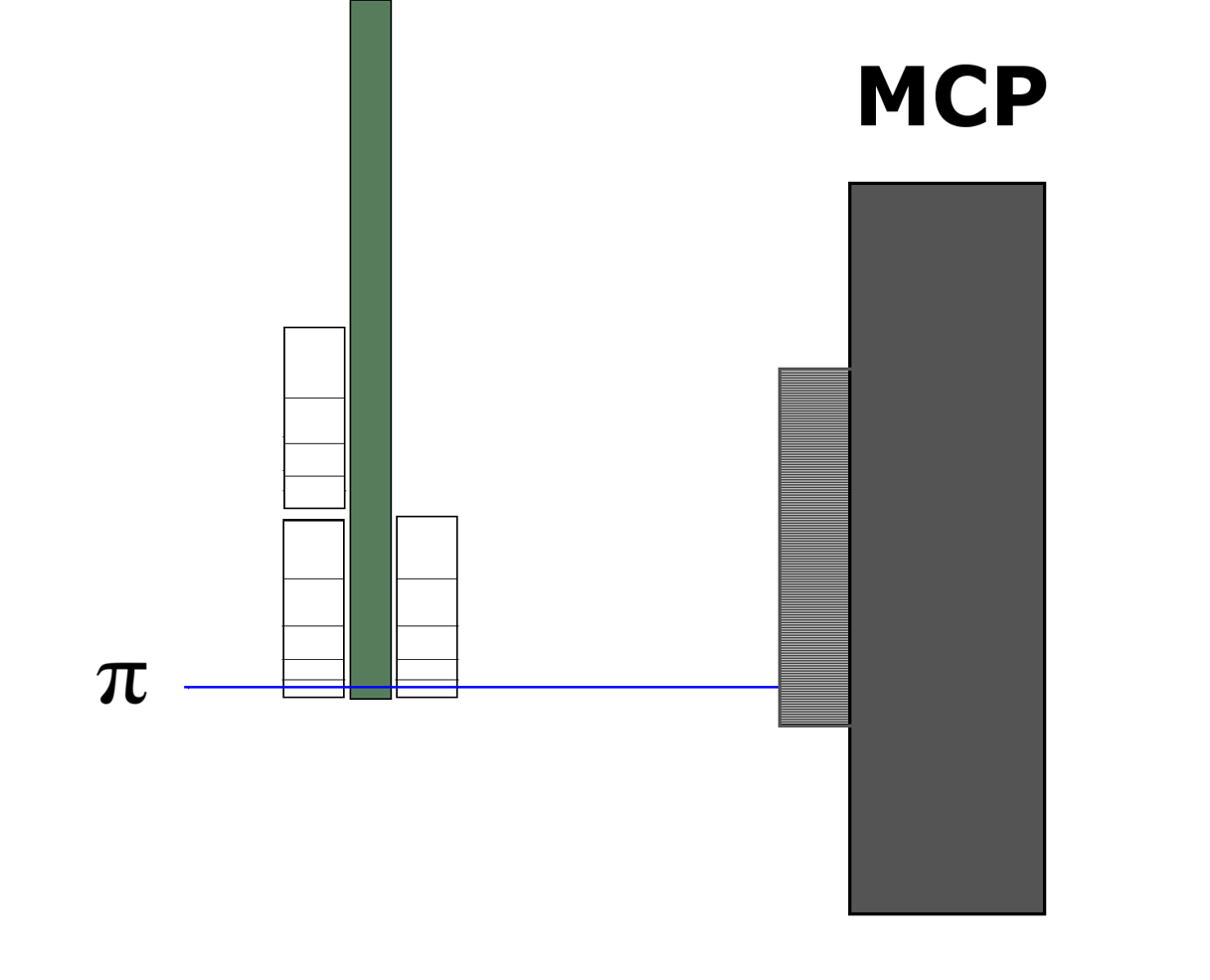}\\
\includegraphics[width=.9\linewidth,]{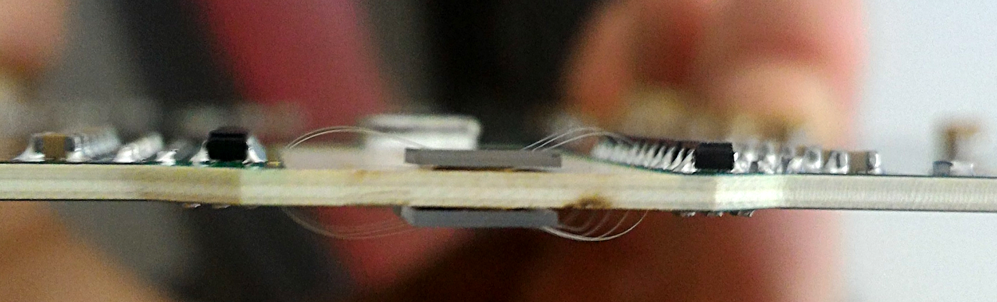}
\caption{Top: sketch of the test beam setup showing the prototype board (hosting the DD and a single diamond) and a Micro Channel Plate. Bottom: a picture of the edge of the prototype, the connections of the DD electrodes on the same amplifier are clearly visible.}
\label{connection-scheme}
\end{center}
\end{figure}

Figure~\ref{connection-scheme} shows the actual realization of the connections and a sketch of the setup used for the test beam measurements.

A Micro Channel Plate (MCP) PLANACON$^{TM}$ 85011-501~\footnote{ PLANACON$^{TM}$ PHOTOMULTIPLIER TUBE ASSEMBLY 85011-501 from BURLE.}  detector used in previous measurements was aligned with the detector and used as a time reference as indicated in Figure~\ref{connection-scheme}.
From measurements performed with an Ultra Fast Silicon Detector (UFSD)~\cite{UFSD-TOTEM}, the time precision  expected from the MCP is of the order of 40 ps .

\begin{figure}[ht]
 \begin{center}
\includegraphics[width=.99\linewidth]{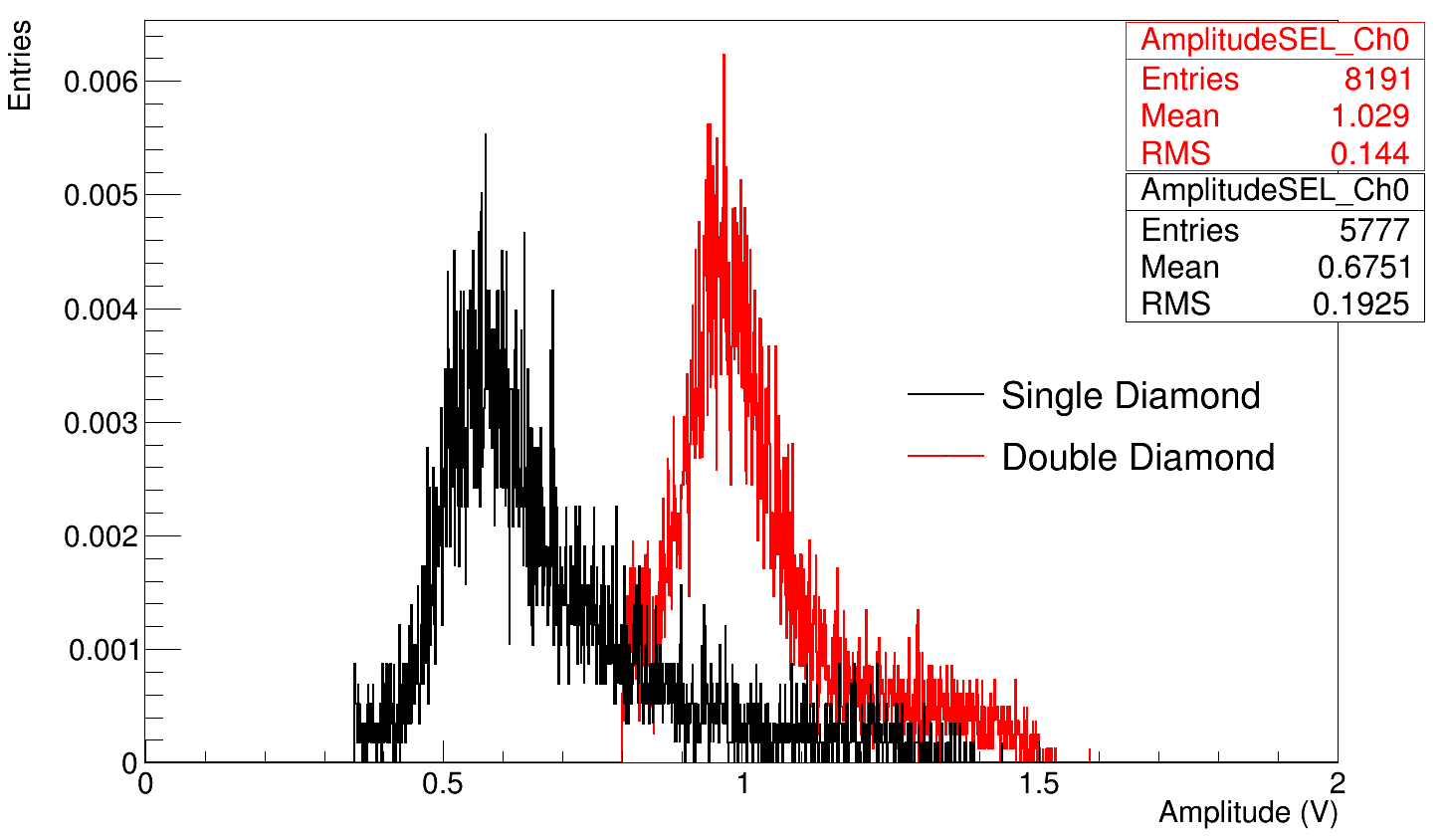}
\caption{Comparison of the measured pulse height for a detector with one 500 $\mu$m crystal (black), and one which mounts two 500 $\mu$m crystals mounted back to back on the same amplifier (red).}
\label{signal-comp}
\end{center}
\end{figure}

The flux of particles traversing the detector was  about 1 KHz/mm$^2$/spill.
Signals are recorded with an Agilent DSO9254A oscilloscope (8 bits, 20 GSa/s)  and an off line constant fraction discriminator technique is used to measure the time difference between the MCP and the diamond signal.

The amplification chain working point has not been re-optimized for best time precision.  
In this first measurement to compare the performance of the two connection schemes implemented on the same board the gain of the DD preamplifier has been reduced to avoid saturation of the signal.

Figure~\ref{signal-comp} shows the comparison of the signal maximum amplitude as measured for a single and a double diamond mounted on the same hybrid.
The average values are 0.67\,V for the single diamond and 1.03 for the double.
The DD rise time was $\sim1.5$\,ns, which is compatible with the single crystal one.

\begin{figure}[ht]
\centering
\includegraphics[width=.89\linewidth]{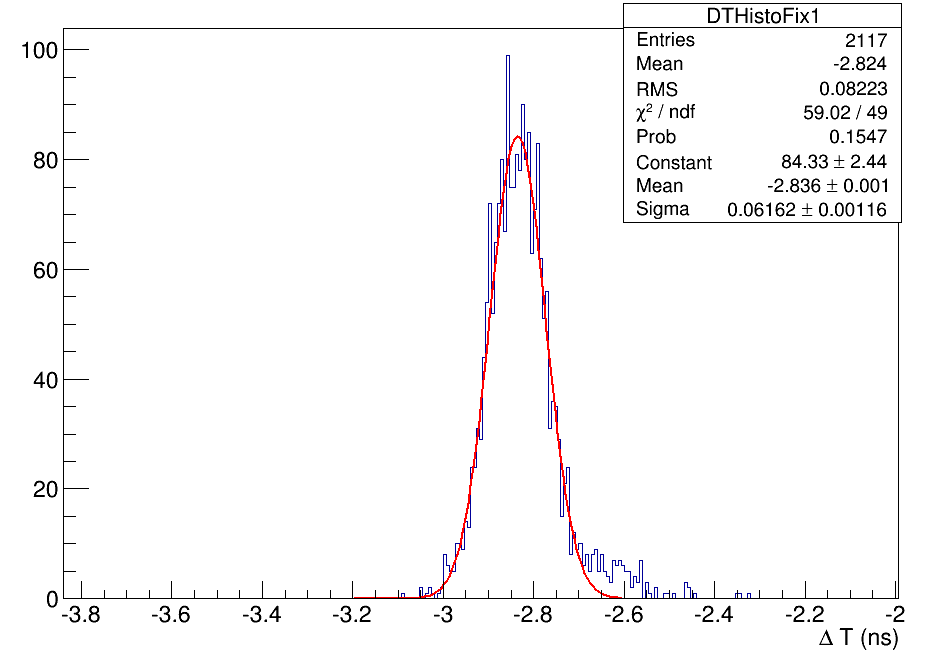}
\caption{Time precision of a double diamond detector.}
\label{time-res}
\end{figure}

Figure~\ref{time-res} shows the arrival time difference between the double diamond and the MCP.
The standard deviation of the Gaussian distribution is 62 ps which means, assuming the MCP time precision of 40\,ps, a time precision of the double diamond of less than 50\,ps.
The small non-Gaussian tail at larger $\Delta T$ values contains $\sim$5\% of the events of the plot.
Therefore an improvement of the time precision of about 1.6-1.7 is obtained with respect to the single diamond, which was measured to be $\sim$80\,ps for a sensor with the same capacitance.

We expected that a slight offset in time in the generation of the charge from the two sensors may degrade the timing capability of the measurement.
In fact due to the position of the amplifier on the side of the board, a small degradation is observed for measurements performed with the beam traversing the two sensors in the opposite directions.
This can be easily improved by modifying the way the double diamond sandwich is realized.
The perfect alignment of the sensors on each side of the board is also of great importance to ensure a good measurement efficiency. 
This is one of the reason why the DD time precision measured here is not exactly equal to half of the single diamond time resolution, but better anyway than the one expected for measurements performed simultaneously on two independent diamonds.

\section{Conclusions}

The TOTEM TOF diamond detectors for the upgrade of the vertical RPs has been operated successfully in the LHC.
Here we have presented  a way to improve the timing precision by reading simultaneously two sensors with the same amplification chain that had been developed for the TOTEM upgrade.


A time precision improvement of a factor  1.6-1.7 has been measured for a double diamond setup when compared to the single plane detector used in  the TOTEM upgrade.

\section*{Acknowledgments}

We thank Florentina Manolescu and Jan Mcgill for the realization of the unusual bonding of the sensors.
Support for some of us to travel to CERN for the beam tests was provided by AIDA-2020-CERN-TB-2016-11.
This work was supported by the institutions listed on the front page and also by the project LM2015058 from Czech Ministry of Education Youth and Sports. 

\bibliography{Diam-2016-DD-J}




%
%
\bibliographystyle{unsrt}

\end{document}